\documentstyle[twocolumn,aps,psfig]{revtex}

\begin{document}
\draft


\twocolumn[\hsize\textwidth\columnwidth\hsize\csname
@twocolumnfalse\endcsname

\title{Mass enhancement of two-dimensional electrons in thin-oxide
Si-MOSFET's} 

\author{W. Pan and D.~C. Tsui}
\address{
Department of Electrical Engineering,
Princeton University,
Princeton, New Jersey 08544}
 
\author{B.~L. Draper}
\address{
Sandia National Laboratories,
Albuquerque, New Mexico 87185}
 
\date{\today}
\maketitle
 
\begin{abstract}
We wish to report in this paper a study of the effective mass ($m^*$) in
thin-oxide Si-metal-oxide-semiconductor field-effect-transistors, 
using the temperature dependence of the Shubnikov-de
Haas (SdH) effect and following the methodology developed by J.L. Smith and
P.J. Stiles, Phys.\ Rev.\ Lett. {\bf 29}, 102 (1972). 
We find that in the thin oxide limit, when
the oxide thickness $d_{ox}$ is smaller than the average
two-dimensional electron-electron separation $r$, 
$m^*$ is still enhanced and the
enhancement can be described by $m^*/m_B = 0.815 +
0.23(r/d_{ox})$, where $m_B = 0.195 m_e$ is the bulk electron mass,
$m_e$ the free electron mass.
At $n_s = 6 \times 10^{11}$/cm$^2$, for example,
$m^* \simeq 0.25 m_e$, an enhancement doubles that previously reported
by Smith and Stiles. Our result shows that the interaction between 
electrons in the semiconductor and the neutralizing positive charges on
the metallic gate electrode
is important for mass enhancement. We also studied the magnetic-field
orientation dependence of the SdH effect and deduced a value of $3.0 \pm
0.5$ for
the effective $g$ factor in our thin oxide samples. 
\end{abstract}
\vskip2pc]

Over two and a half decades ago, Smith and Stiles\cite{stiles:prl72}
reported their observation of effective-mass enhancement in
the two-dimensional electron system (2DES) 
in silicon metal-oxide-semiconductor field-effect transistors
(Si-MOSFET's). Following an
earlier experiment by Fang and Stiles\cite{fang:prb68} on effective
$g$ factor enhancement, they were able to take advantage of the
continuous density tunability of the device and carefully investigated
the 2D electron density dependence of mass enhancement. Their
finding that the effective mass $m^*$, in the unit of free electron
mass $m_e$, continuously increases from $\sim$ 0.21 to 0.225 for
electron density ($n_s$) decreasing from $\sim 3 \times 10^{12}$/cm$^2$ to
$0.6 \times 10^{12}$/cm$^2$ demonstrates unambiguously the
electron-electron ($e-e$) interaction origin of the enhancement. Their
experimental result was soon confirmed by theory and stimulated a
great deal of theoretical interest in many-body effects in low
dimensional electron systems. Subsequently, Fang, Fowler, and
Hartstein\cite{fang:prb77} also investigated the influence of oxide
charge and interface states on effective-mass measurement. 

One question that was not addressed in these pioneering works is how
the oxide thickness influences the mass enhancement. In particular, as
the oxide thickness decreases, screening of the $e-e$ interaction by
the metallic gate will become important. 
In the limit when the average $e-e$ separation is larger
than the oxide thickness and the gate screening prevails, the
$e-e$ interaction induced mass enhancement can be expected to diminish. 
In other words, 
if there is no other
mechanism for enhancement, the mass as a function of $n_s$ should
show a decrease with decreasing $n_s$ in this low density limit.

We wish to report in this paper a study of the effective mass in
thin-oxide Si-MOSFET's, using the temperature dependence of the Shubnikov-de
Haas (SdH) effect and following the methodology developed by Smith and
Stiles \cite{stiles:prl72}. We find that in the low-density limit, when
the oxide thickness $d_{ox}$ is smaller than the average 2D
$e-e$ separation $r$, defined by $r = 1/\sqrt{n_s}$, $m^*$ is still
enhanced and the enhancement can be described by $m^*/m_B = 0.815 +
0.23(r/d_{ox})$, where $m_B$ is the bulk electron mass
equal to $0.195 m_e$. At $n_s = 0.6 \times 10^{12}$/cm$^2$, for
example, $m^* = 0.25 m_e$, an enhancement doubles that previously
reported by Smith and Stiles. We also studied the magnetic-field
orientation dependence of the SdH effect and deduced a value of $3.0 \pm
0.5$ for the
effective $g$ factor in our thin oxide samples.
 
The samples are $n$-type inverted silicon(100) surfaces on
$p$-type substrates with peak mobility 1700 cm$^2$/V~s at 4.2 K.
The thickness of the oxide ($d_{ox}$) is 60 \AA,
The channel length ($L$) is 2 $\mu$m
and the channel width ($W$) 12 $\mu$m. Since $W/L \gg 1$,
the edge effect is not important in our
experiments. 
The leakage current between the gate and the
channel is virtually zero and much smaller than the 
drain-to-source current ($I_{DS}$).
A drain voltage ($V_D$) not larger than 20 $\mu$V is applied throughout
the experiment to ensure that the measurements
are done in the Ohmic regime and there
is no hot-electron effect.
Transconductance, $G_m$ = $dI_{DS}/dV_G$, is measured 
at a fixed magnetic field (here $V_G$ is the gate voltage). 
The magnetic field ($B$) is chosen according 
to the following: (1) It has to be
small enough so that the experiment is done in the 
SdH oscillation regime. (2) The $B$ field has to be sufficiently large
so that the SdH oscillations are clearly observed.
The experiments are performed in a $^3He$ system, with a base
temperature ($T$) of 0.3 K. 
The temperature uncertainty generally
is smaller than $\pm0.005$ K.
The measurements
are done by employing a standard low-frequency lock-in
technique, typically at 7 Hz. 
The results are reproducible from run to run after the same cool down.
As many as ten samples of the same type are studied. All the samples
give the same results within our experimental error.

In Fig.~1, we plot $G_m$ as a function of $V_G$ for one of
the samples (sample A). 
The SdH oscillations are observed in electron densities down to
$0.6 \times 10^{12}$/cm$^2$. 
$G_m$ has been measured at many temperatures, and here three
different temperature traces are shown to illustrate the temperature
dependence of the
oscillation amplitude. After the nonoscillatory
background is subtracted, the oscillations are found to be
sinusoidal with a single period,
which proves that all
electrons occupy one subband. The period of the oscillations is the
change in $V_G$ which produces a change in $n_s$ equal to the number
of electrons needed to fill a Landau level. This number is
$4eB/h$ (here, $e$ is the electronic charge, $h$ is Planck's
constant). Consequently, the minima in $G_m$ vs. $V_G$
occur whenever $V_G$ satisfies

\begin{equation}
V_G = N(4e^2/hC_{ox})B + V_{th},
\label{linearT}
\end{equation}
where $V_{th}$ is the threshold voltage, $C_{ox}$ is the oxide
capacitance, and $N$ is an integer. Thus, if we label the
minima (maxima) in $G_m$ vs $V_G$ from left to right by
integer (half-integer) number $N$, 
a plot of $N$ against $V_G$ at
which the minima (maxima) occur should yield a straight line.    
The inset of Fig.~1 shows such a plot, where the
filled dots are the minima of the oscillations and the open dots the
maxima.  
It is clear that the data points lie on a straight line and $N$ follows
a strictly linear dependence on $V_G$. The rate of change of $n_s$
with respect to $V_G$, i.e., $dn_s/dV_G = C_{ox}/e$, obtained from the
slope of the straight line is $3.4 \times 10^{12}$ /cm$^2$~V, in good
agreement with that calculated from the oxide thickness.
The intercept with the $x$ axis at $N = 0$ 
gives $V_{th} = 0.57$ V for this sample. 

The SdH formalism is used to derive the effective mass 
\cite{stiles:prl72}. The absolute amplitude ($A$) of the
SdH oscillations is obtained by subtracting the non-oscillatory 
background and drawing the
envelope on the oscillatory part of the $G_m$ traces. At each
$V_G$, a set of amplitudes is then obtained from a set of temperature
dependence data. The amplitude is 
fitted by a non-linear least-squares
technique according to the equation

\begin{equation}
A \sim \frac{\xi}{sinh(\xi)},
\label{linearT}
\end{equation}
where $\xi = 2\pi^2k_{B}T/\hbar\omega_{c} = 2\pi^2k_{B}Tm^*/e\hbar B$,
$\omega_{c}$ is cyclotron frequency, and $m^*$ is the effective
mass. All other symbols have their usual meanings. 
In the fitting, it has been assumed 
that the relaxation time $\tau$ is independent of 
temperature\cite{stiles:prl72}.

In Fig.~2(a), a plot of $m^*$ vs $n_{s}$ is shown
for two samples. The solid symbols represent the
results from sample A,
and the open ones from sample B\cite{error}.
The effective mass shows an
unexpectedly strong increase at low densities, up to 
0.25$m_e$ at $0.6 \times
10^{12}$/cm$^{2}$, even larger
than what was reported by Smith and Stiles (0.225$m_e$ at the same density) 
\cite{stiles:prl72},
which is shown as the dashed line.  
The data cover the density range from $0.6 \times 10^{12}$/cm$^2$
to $2.0 \times 10^{12}$/cm$^{2}$, corresponding to an average $e-e$
separation varying from 
$r$ = 130 \AA~to 70 \AA.
For $n_s < 0.6 \times 10^{12}$/cm$^2$, 
the absolute amplitude of the SdH oscillations is quite small while the
nonoscillatory background is relatively large and steep. This fact
makes the determination of the oscillation amplitude and the data
fitting unreliable.

As we have mentioned above, in a thin-oxide Si-MOSFET, the effective
mass is expected to decrease with decreasing density in the low
electron density limit where the screening of the $e-e$ interaction by
the metallic gate prevails. 
The observed strong mass
enhancement, therefore,  must originate from another mechanism, which becomes 
important in the low $n_s$ limit. 
First, we want to point out that unavoidable charges 
in the oxide can not be the origin. The electron mass is known to
show an opposite $n_s$ dependence when a high
concentration of charges is incorporated in 
the oxide\cite{fang:prb77}.
Also, we recall that a MOSFET
is simply a capacitor with the metallic gate and 
the semiconductor its two plates. 
The application of a gate voltage
changes the amount of charge on the capacitor plates. 
With a positive $V_G$ applied, electrons are transferred from the
metallic gate electrode onto the silicon and are stored in the 2DES 
at the silicon
surface. The metallic gate electrode, in turn, attains a sheet of
positive charges,
which can be conveniently 
viewed as mobile holes. Their influence 
on the dynamics of the 2DES is through the Coulomb screening
\cite{ando:rmp82},
the strength of which is controlled by $r/d_{ox}$.
In Fig.~2(b), $m^*$ is replotted as a
function of $r/d_{ox}$. 
We find that $m^*$ shows a linear dependence on $r/d_{ox}$. 
In fact, within experimental error, all our data can be fitted by 
$m^*/m_B = 0.815 + 0.23(r/d_{ox})$.
This empirical fit obviously breaks down in the $r \ll d_{ox}$ limit,
where $m^*$ is known to equal the bulk band mass $m_B = 0.195m_e$.
In the $r/d_{ox} \to \infty$ limit, binding of the 2D electrons and
the neutralizing holes on the metallic gate electrode is expected to
form a 2D dipole gas and $m^*$, then, should be the mass of the
electric dipole.
  
The mass enhancement due to the presence of the 
mobile holes on the gate electrode can be viewed as resulting from 
the Coulomb drag 
effect\cite{price:physicab83,solomon:prl89,gramila:prl91,sivan:prl92} 
--- an interlayer Coulomb coupling between
the electron layer in the semiconductor and 
the hole layer on the metallic gate electrode.
In such a two layer system, when a current is driven through the electron
layer, the interlayer $e-h$ interaction creates a
frictional drag force that can  modify the electron self-energy
and change the effective mass. Several recent experiments
\cite{solomon:prl89,gramila:prl91,sivan:prl92} measured the
interlayer scattering rate in bilayer parallel electron and parallel
hole systems. Sivan, Soloman, and Shtrikman \cite{sivan:prl92} 
studied the bilayer $e-h$ system in a
GaAs/Al$_{1-x}$Ga$_x$As heterostructure and found that the scattering rate
increases with decreasing $n_s$, similar to the $n_s$ dependence of
the mass enhancement observed in our experiment.

The magnetic-field orientation ($\theta$) dependence of the SdH effect is also
studied, from which an effective 
$g$ factor is deduced by the coincidence method developed by Fang and
Stiles\cite{fang:prb68}. 
Figure 3 shows two $G_m$ traces at different tilt angles, $\theta = 0^{\circ}$ and
$\theta = 68.0^{\circ}$. The total $B$ field is different for the two cases ($B$ =
3.75 and 10.0 T respectively), but the perpendicular $B$ field ($B_{\bot} =
3.75$ T) is kept the same.
The phase of the two traces is roughly $180^{\circ}$ different, or reversed
\cite{discrepancy}. For
example, the minimum at $V_G = 1.02$ V in the $\theta = 0^{\circ}$ trace
becomes a maximum in the $\theta = 68.0^{\circ}$ trace. 
The phase reversal can be understood from the fact that, while the Landau
level separation ($\hbar \omega_c$) depends only on $B_{\bot}$, the spin
splitting ($g\mu_BB$) depends on the total $B$ field. As illustrated in the
inset of Fig.~3, at $\theta = 0^{\circ}$, two spin-degenerate 
$N$th, ($N+1$)th Landau levels
give rise to the maxima at $V_G = 0.96$ and $1.09$ V.  
By increasing $\theta$,
the spin degeneracy of Landau levels
is lifted.  The spin-up $N$th 
Landau level goes up while 
the spin-down ($N+1$)th Landau level 
moves down.
At $\theta = 68.0^{\circ}$, $g\mu_BB =
\hbar\omega_c$ and the two levels cross to form the 
spin-degenerate Landau level giving rise to the maximum at $V_G = 1.02$ V.
The effective
$g$ factor can be estimated 
from $g\mu_BB =
\hbar\omega_c$ and we found for our samples $g = 3.0 \pm 0.5$.
The large error bar is mainly due to the large Landau-level broadening
and nonzero spin splitting at zero angle.
Within this large experimental error,  
$g$ factor shows no density
dependence. Finally, similar large $g$ factor enhancement was observed
by Van Campen \cite{campen:thesis94} who studied the magnetoconductance 
in high-$B$ fields where the spin splitting is resolved. By fitting the line
width of the magnetoconductance oscillations, he obtained a $g$ factor
from 2.5 to 3.6 at different densities in his sample 
with $d_{ox} = 44$ \AA~and peak
mobility 6300 cm$^2$/V~s.  

In conclusion, we have observed an unexpected mass enhancement in the
low-density limit in
thin-oxide Si-MOSFET's. The enhancement is attributed to the presence
of the positive neutralizing charges on the metallic gate electrode at a
distance smaller than the average $e-e$ separation. 
The effective $g$ factor is also measured.
Its value, $3.0 \pm 0.5$, is larger than the bulk value and shows no
dependence on the electron density within our experimental resolution. 

We acknowledge help from J.P Lu and useful discussions with Harry
Weaver and Kun Yang.
The work done at Sandia was supported by the U.S. Dept. of Energy (DOE).
Sandia National Laboratories is operated for DOE by Sandia
Corporation, a Lockheed Martin Company, under Contract No.
DE-AC04-94AL85000. 
The work at Princeton University was supported by the NSF.

\begin{figure}[htbp]
\vspace{8.2cm}
\centerline{\psfig{figure=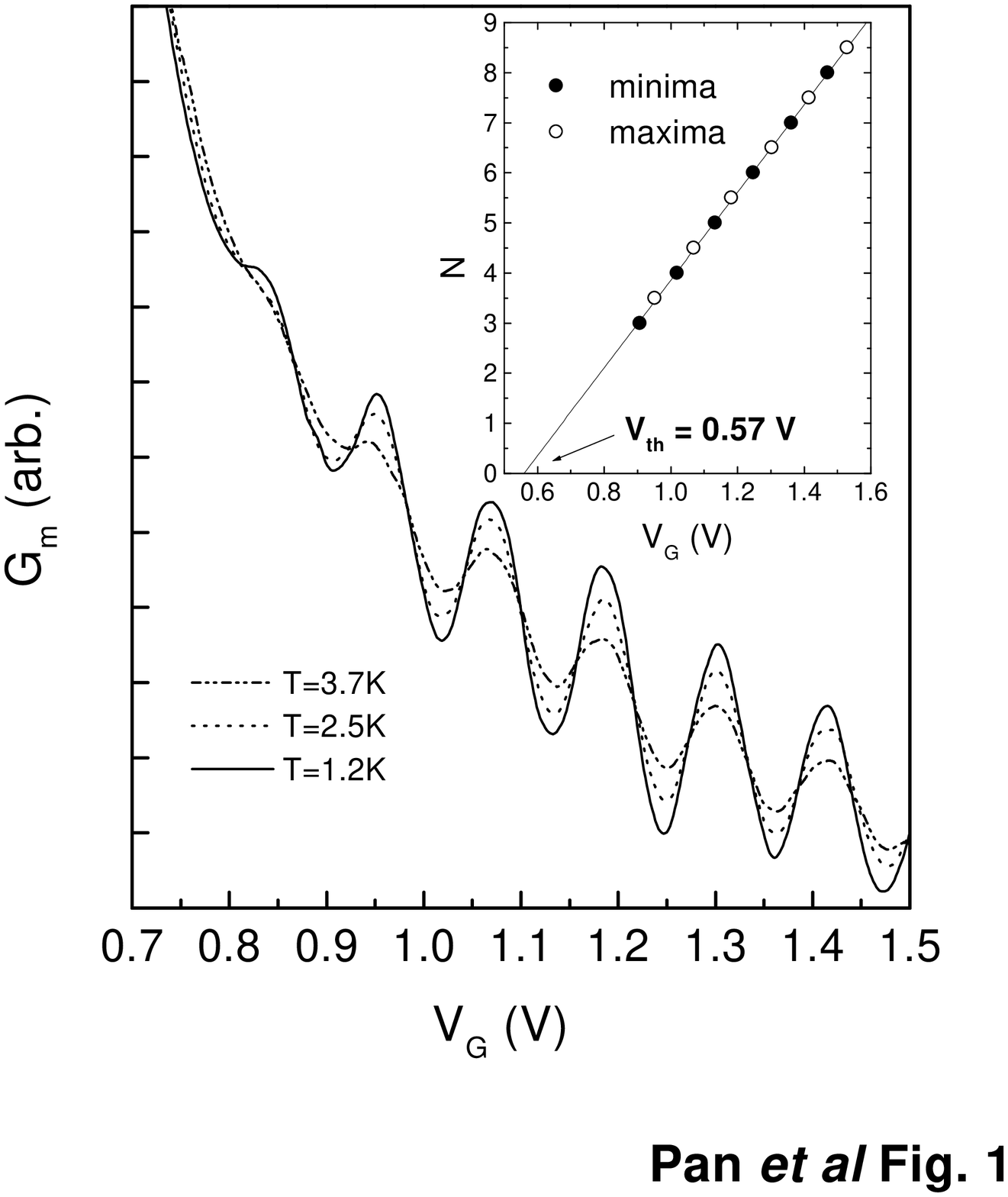,width=7.5cm,angle=0}}
\vspace{0.2cm}
\caption{
The temperature dependence of $G_m$ for
sample A.
Three temperature traces are shown at $T$ = 1.2, 2.5, 3.7 K.
The inset shows $N$ vs $V_G$ at which the minima (maxima) of $G_m$
occur.
The close dots are
the minima, the open ones the maxima. The straight
line is a linear fit to the data. $V_{th}$ is the threshold voltage.
}
\vspace{0.2cm}
\end{figure}

\begin{figure}[htbp]
\vspace{8.5cm}
\centerline{\psfig{figure=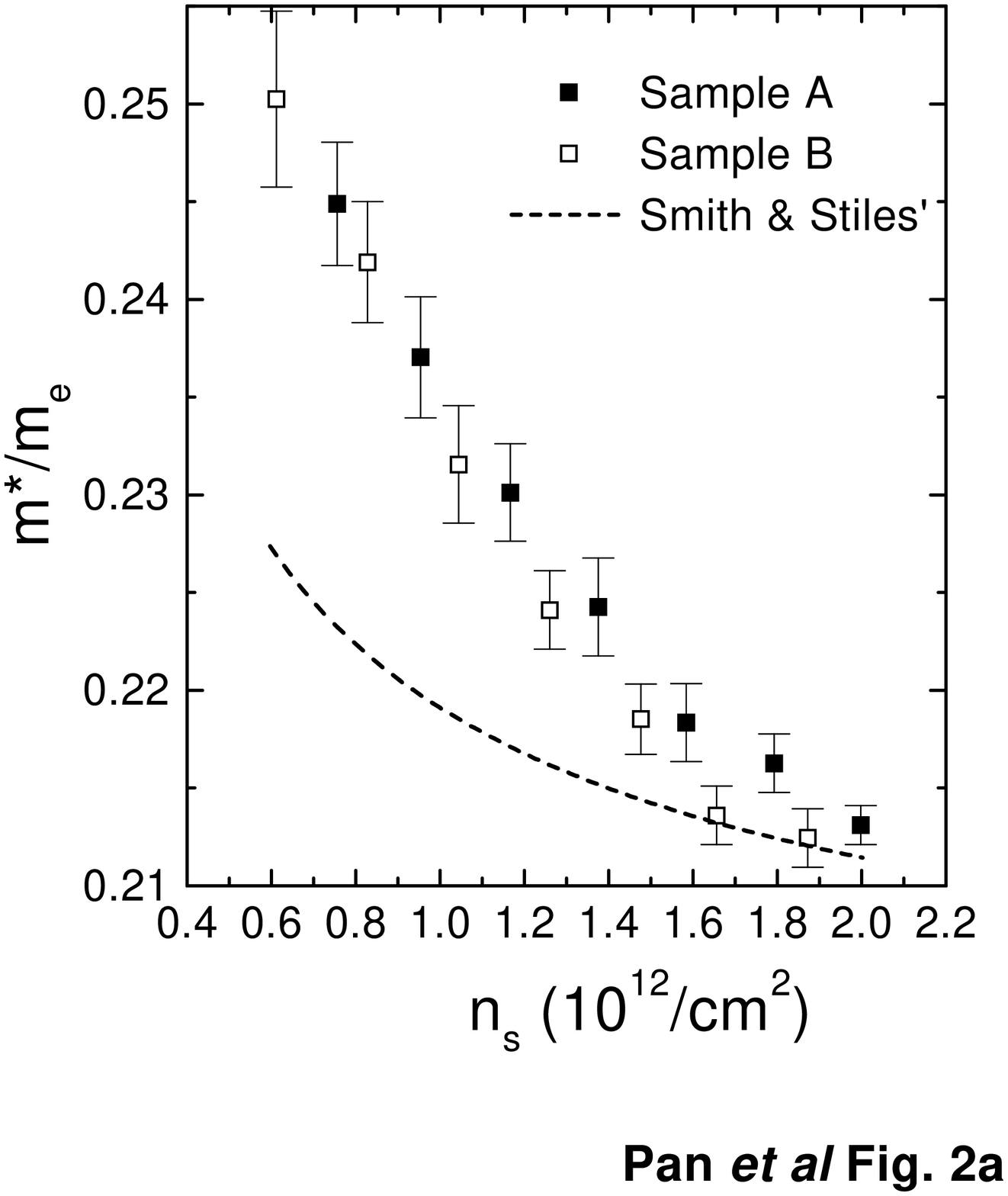,width=7.5cm,angle=0}}
\vspace{0.2cm}
\centerline{\psfig{figure=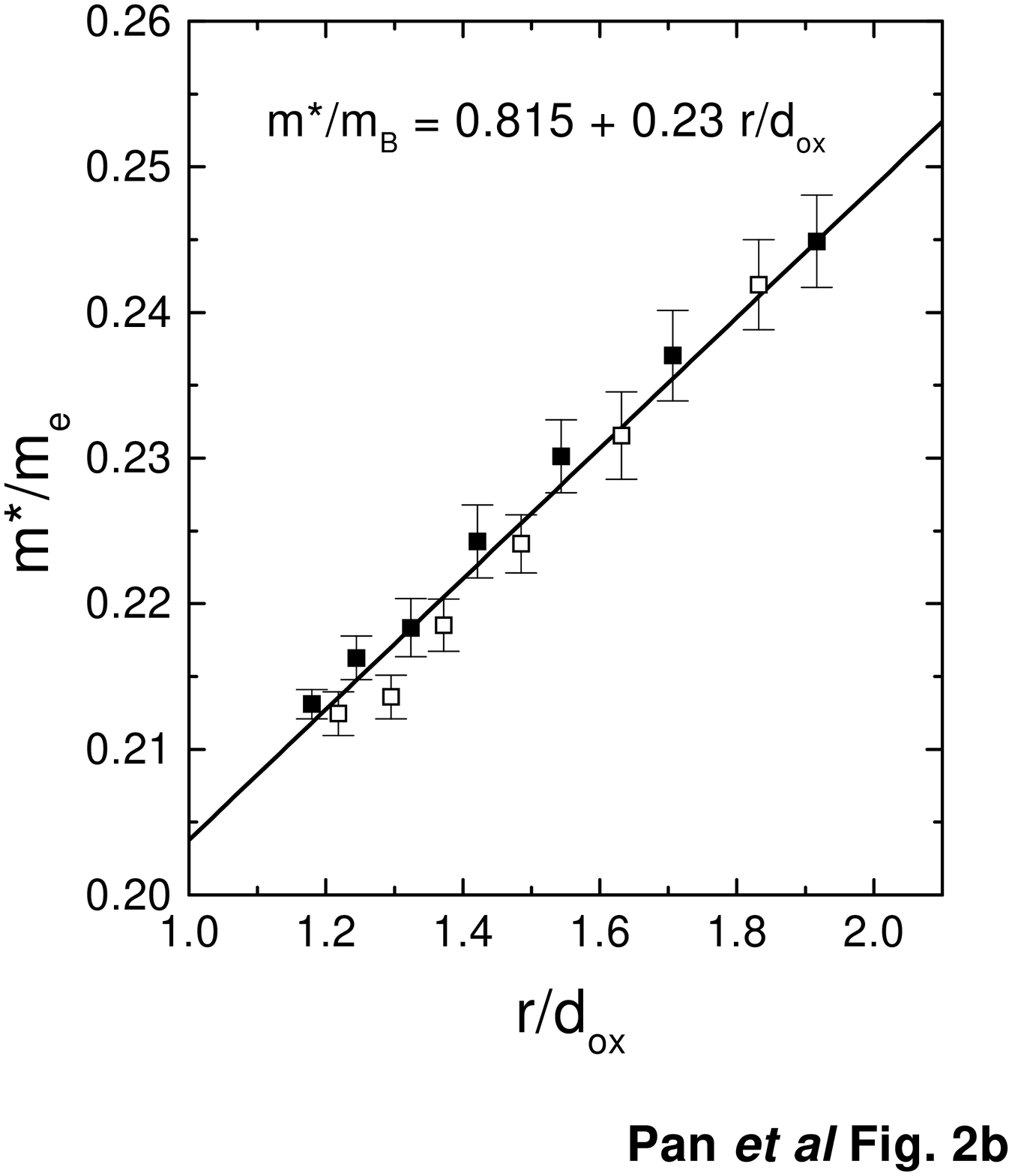,width=7.5cm,angle=0}}
\vspace{0.2cm}
\caption{
(a) $m^*$ vs $n_s$ for two samples,
sample A (the solid symbols) and sample B (the open symbols).
The dash line represents
the previous result from Smith and Stiles.
(b) The replot of $m^*$ as a function of
$r/d_{ox}$. $m^*$ shows a linear
dependence on $r/d_{ox}$.
}
\vspace{0.2cm}
\end{figure}

\begin{figure}[htbp]
\vspace{-0.5cm}
\centerline{\psfig{figure=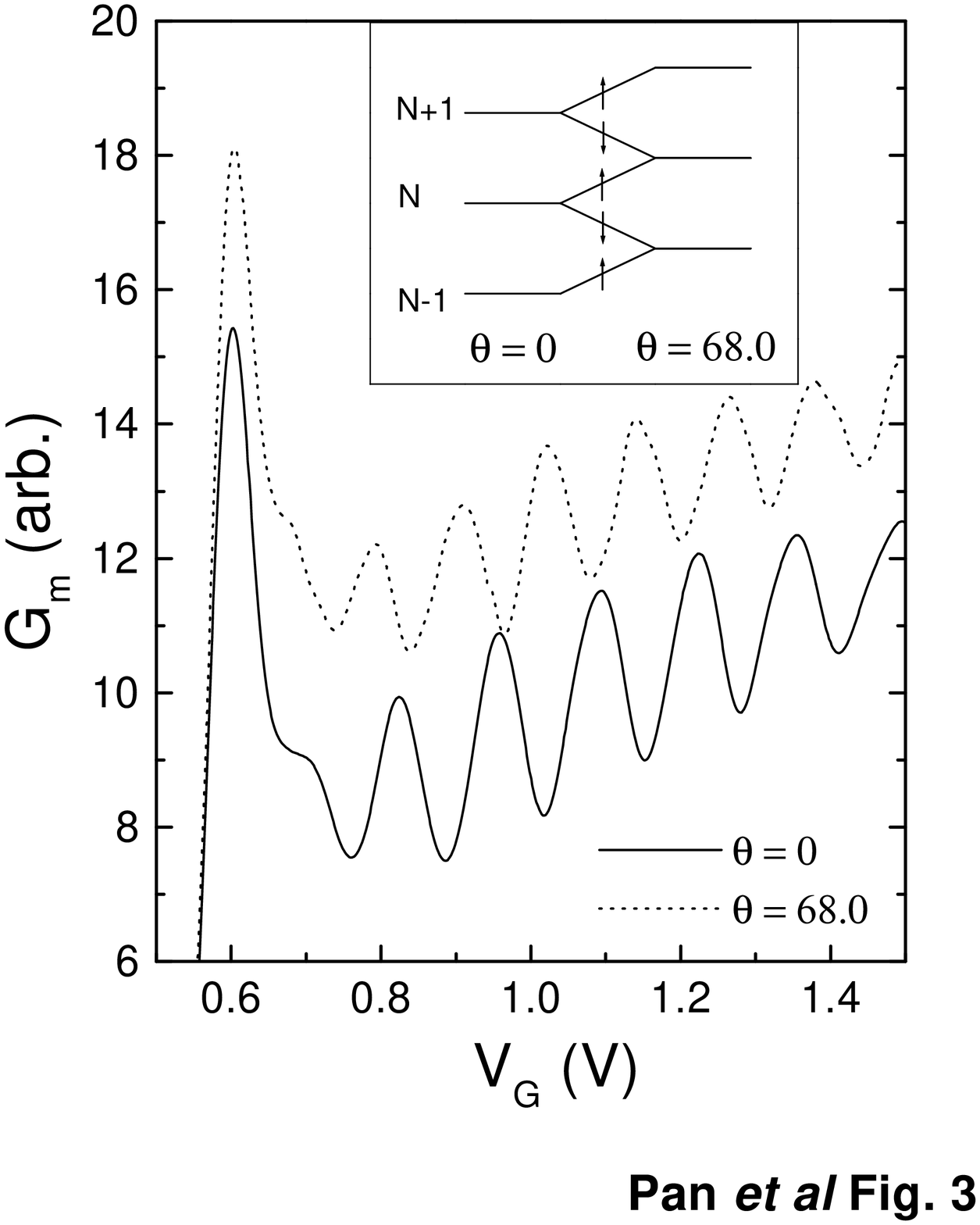,width=7.5cm,angle=0}}
\vspace{0.2cm}
\caption{
$G_m$ vs $V_G$ for $B=3.75$ T at
$\theta = 0^{\circ}$ and $B=10.0$ T at $\theta=68.0^{\circ}$.
$B_{\bot}$ is the same for the two cases.
The upper trace is shifted
vertically by a factor of 3 units. The
phase reversal is clearly seen for the two traces. The inset shows the
evolution of Landau levels with the tilt angle.
}
\vspace{0.2cm}
\end{figure}


\begin{references}
\bibitem{error}
	The error bar represents one standard
statistical deviation from data fitting. The error at higher density
is smaller. The trend is real.
\bibitem{discrepancy}
   The period of SdH oscillations differs by 
approximately 10-15~$\%$ at the two angles
   in Fig.~3.
   There are several sources contributing to the difference.
   The first and the most important one is the uncertainty of the
   tilted angle, about $2^{\circ}$ from the
   uncontrollable backlash of the gear in our probe.
   The second is due to the large Landau-level broadening and nonzero
  spin splitting at zero angle in our samples,
which is also responsible for the
large error bar in the $g$ factor. Under this circumstance, the maxima of
the SdH
oscillations do not necessarily occur exactly at $g\mu_{B}B = \hbar
\omega_c$.
Instead, they can happen in a certain range of tilted angle.
Finally, deformation of
the electron wave function due to the large in-plane $B$ field may also
change the oscillation period for the two tilted angles with the same
perpendicular
$B$ field.
\end{references}
\end{document}